# Effect of Added Salts on the Interfacial Dynamics of Monovalent Metal Ions and Model Water-Soluble Polymers

Soumik Ghosh[1, 2], Jack F. Douglas[1, *], and Francis W. Starr[2]

[1]Materials Science and Engineering Division, National Institute of Standards and Technology,

Gaithersburg, Maryland 20899, USA

[2]Department of Physics, Wesleyan University, Middletown, Connecticut 06459-0155, USA

## ABSTRACT

Traditional theories of electrolyte and polyelectrolyte solutions overlook ion and polymer hydration, yielding an overly idealized perspective of the thermodynamic and dynamic properties of these solutions. In particular, the propensity of certain ions to increase or decrease the water diffusion coefficient at low solution concentrations, is poorly understood, a phenomenon that correlates strongly with the Hofmeister series describing the influence of salts on the solubility and self-assembly of synthetic and biomacromolecules as well as colloidal particles. These ion-specific effects have profound ramifications in diverse biological, medical science and technological applications. We address the general problem of understanding ion and uncharged water-soluble polymer hydration through molecular dynamics (MD) simulations of representative water-soluble polymers, polyacrylamide (PAM) and polyethylene oxide (PEO), in aqueous salt solutions. Careful attention is given to the capacity of our model to reproduce observed trends of added monovalent salts on the diffusion coefficient of water. Reproducing these ion-specific trends in the mobility of water has been a recurrent challenge in prior MD simulations of aqueous solutions based on a physically realistic model of water. Our work on salt solutions represents an important advance on which we build our modeling of aqueous polymer solutions with added salts. Simulations of our model polymers reveal an extended nanoscale dynamic hydration layer having a scale on the order of one nanometer (nm) in which the mobility is perturbed from its bulk value, and the mobility gradient in this layer appears to obey a near universal functional form.

Corresponding authors: soumik.ghosh@nist.gov; jack.douglas@nist.gov

*Emeritus

## I. Introduction

Understanding the conformational and dynamical properties of polymers in aqueous environments is important for their application in diverse chemical, biological, and industrial fields[1–3]. Among the non-toxic polymers that are soluble in water, poly(ethylene-oxide) (PEO) and poly(acrylamide) (PAM) have gained significant interest due to their high solubility and diverse uses in drug delivery[4] and water treatment[5]. The interactions of these polymers with water molecules greatly affect many of their equilibrium and non-equilibrium properties due to the formation of complex hydration structures around the polymer backbone[6]. The behavior of water within these hydration layers plays an essential role in determining the flexibility of the polymer backbone, segmental movement, and overall solution characteristics such as viscosity and diffusion[7,8].

When salts are added to aqueous solutions, hydration shells thickness form. These hydrated ions are dynamic complexes, which alter both the thermodynamics and dynamics of the aqueous solution in an ion-specific fashion, has long resisted a general explanation. In a commonly mentioned intuitive model by Frank and Wen[9], ions are imagined to promote or disrupt the water hydrogen-bonding network leading to "structure making" or "kosmotropic" ions, and "structure breaking" or "chaotropic" ions[10] in their figurative language[10]. Operationally, kosmotropic ions are conventionally defined experimentally as those that increase the shear viscosity of the aqueous salt solution while chaotropic ions corresponding decrease the shear viscosity where the leading viscosity virial coefficient (viscosity $B$ coefficient[11]) serves to quantify the effect for any given ion. Anions and cations of the salt contribute nearly interpedently to $B$, a simplification that allows the determination of tables of $B$ coefficients for a diverse range of ions [12]. This phenomenon leads to corresponding changes in the water self-diffusion coefficient $D$, where kosmotropic ions decrease $D$, and chaotropic ions increase $D$. Thus, chaotropic ions lead to changes in the aqueous salt solution where the virial coefficient for the water diffusion coefficient ($\approx$-$B$) is positive, at least for temperatures near ambient conditions where the Stokes-Einstein relation relating the shear viscosity of the solution to the diffusion coefficient of the solvent is approximately valid[13]. The question of whether any actual change in the structure of water occurs upon adding salts has long been debated, but scattering evidence consistent with this structural interpretation of kosmotropic and chaotropic ions indicates that naïve Frank and Wen model is highly questionable[14,15]. Most of the classical molecular dynamics (MD) simulations models, aimed at modeling even the "simple system" of salt solutions at ambient temperature and pressure, have repeatedly failed, even qualitatively, to correctly describe the experimentally measured changes of the dynamics and thermodynamics of these solutions[11,16,17]. In particular, past classical MD simulations based on a wide range of potential models, both polarizable and non-polarizable, could not even account for the evidence of chaotropic ions[11]. Some recent works based on the TIP4P/ 2005 model have demonstrated some success in reproducing the effect of chaotropic ions based on the artificial assumption that the ions in solution have various fractions of the bare ion charge [18]. However, this type of "partial charge" model for the ions has been criticized from a theoretical standpoint, and at a semi-empirical level this approach also has the downside of degrading the quality of electrolyte solution thermodynamic property estimates[19]. We have thus not pursued this approach.

Given the success of the TIP4P/2005 water model to provide a reliable description of many of the transport and thermodynamic properties of real water [20], it is perhaps surprising that simulations of aqueous salt solution properties using TIP4P/2005 have not been more successful. Apparently, these properties are highly sensitive to how the potential parameters governing the ion-water interactions are chosen. Here, we focus on TIP4P/2005 as a suitable foundation for our modeling of aqueous solutions, and we explore ion-water potential parameters from prior literature that might provide an accurate description of the electrolyte solutions. Coarse-grained models of aqueous solutions[21,22] are valuable to provide insights for problems where the interactions are complex and this type of modeling has provided insights into the origin of chaotropic ions. in terms of how the ions influence the cohesive interaction of the solution[21,23]. A key feature of coarse-grained models of aqueous salt solutions is the use of ion solvation energy to determine the ion-water interaction potential parameters. Convenient for our purposes, Li et al.[24] utilized this approach to estimate ion-ion and ion-water interaction potentials, although these authors did not consider the impact of their potential parameters on the transport properties of aqueous slat solutions. We find that the potential parameters derived by Li et al. can well-reproduce basic trends in water mobility for a range of ions ranging from kosmotropic to chaotropic. Notably, this approach does not require invoking unphysical partial charges and we also do not utilize approximations of previous coarse-grained calculations of aqueous salt solutions in which the dipolar nature of the solvent was completely neglected[21,22]. We consider this to be a potentially important step forward in the classical MD modelling of aqueous salt solutions in a physically realistic fashion.

It is notable that first principal quantum calculations of aqueous water solutions have been able to reproduce the effects of chaotropic ions[25], at least qualitatively. However, the system sizes accessible by these methods are small due to the computational expense of density functional theory (DFT) methods, obviating the need for more accurate classical models of aqueous salt solutions. Alternatively, machine-learned potentials that are trained from DFT data offer the possibility of quantum accuracy at a fraction of the computational expense[26,27]. Indeed, Avula *et al.*[27] had some success in reproducing experimental trends in solution transport properties of aqueous solutions for two different salts. Here, we consider if traditional approaches might prove equally fruitful and offer more physical insight into the origins of the behavior.

The presence of salts in polymer solutions adds an extra layer of complexity to polymer–water interactions since salts alter polymer conformation and relaxation processes. Such ion-specific effects often lead to changes in the stability of stimuli-responsive polymers. First reported by Franz Hofmeister in 1888[28], different ions, even bearing the same charges, influence the solubility of polymers in an ion-specific fashion. The ranking of ions (separate tables for cations and anions) based on their ability to precipitate long polymer chains is known as the Hofmeister series (HS)[29]. While the kosmotropic ions produce a stabilizing (or, salting-in) effect, chaotropic ions tend to destabilize polymers (salting-out). leading to precipitation. Notably, while one can alternatively define kosmotropic and chaotropic salts based on their effect on viscosity and the diffusion coefficient, this series should not be conflated with the HS; that said, ranking effects based on changes in dynamics strongly correlates with the rankings of the HS. Although such ion-specific effects are ubiquitous, the extent of such phenomenon also depends on different physicochemical properties of the polymer chain, such as, molecular mass, charge density, presence of dipolar interactions[30–33], etc. The presence of salts is known to modulate the balance between hydrophilic and hydrophobic interactions, influencing the microscopic solvation structure. Despite extensive experimental and theoretical studies over several decades, a molecular-level understanding of how different salts affect the coupled dynamics of polymers and surrounding water remains limited.

Our investigation of the effects of monovalent kosmotropic and chaotropic ions on polymer solutions consists of two parts. In the first part of our study, we have investigated changes in structural and dynamical properties of bulk water due to added salts. We simulate salts spanning the kosmotropic to chaotropic range at concentrations up to 1M (molarity), allowing us to test the concentration dependence of water self-diffusion coefficient at room temperature (300 K) and ambient pressure (1 atm). We also examine the water dynamics in the hydration layer of kosmotropic and chaotropic ions. The latter part of our study involves simulation of two different polymers, PEO and PAM, in saline water. We then examine the changes in conformation of those polymer chains due to the addition of the same set of salts at fixed concentration. Finally, we quantify the properties of water molecules in the static and dynamic hydration layer formed around the polymer chains. Curiously, we find that the size of this layer does not depend on the salt used, suggesting a universal nature of the hydration layer, both the spatial extent and the mobility gradient in the layer.

## II. Model and Methods

Polyethylene oxide (PEO) and polyacrylamide (PAM) chains are modeled using an updated all-atom Optimized Potential for Liquid Simulations (OPLS-AA) force field introduced by Jorgensen *et. al*.[34]. The non-bonded interactions between $i^{th}$ and $j^{th}$ particles include a 12-6 Lennard-Jones (LJ) term and a Coulomb term:

$$E_{ij} = f_{ij}\left[\frac{q_i q_j}{r_{ij}} + 4\varepsilon_{ij}\left(\frac{\sigma_{ij}^{12}}{r_{ij}^{12}} - \frac{\sigma_{ij}^{6}}{r_{ij}^{6}}\right)\right] \tag{1}$$

where $q_i$ and $q_j$ are the partial charges on atoms $i$ and $j$, $r_{ij}$ is the distance between the pair, $\epsilon$ and $\sigma$ are potential well depth and diameter, respectively. The term $f_{ij}$ serves as a "switch" to control which pairs interact via non-bonded interactions. Specifically, for intramolecular pairs, separated by larger than three bonds, $f_{ij} = 1.0$, For pairs separated by exactly three bonds, $f_{ij} = 0.5$; for intramolecular pairs separated by less than three bonds, $f_{ij} = 0.0$. We use a geometric mixing rule for the LJ parameters[35], $\sigma_{ij} = \sqrt{\sigma_{ii}\sigma_{jj}}$ and $\varepsilon_{ij} = \sqrt{\varepsilon_{ii}\varepsilon_{jj}}$. Other than non-bonded term, bond stretching, angle bending and torsional energies are also included in the total energy term[36]. Bond lengths of the covalent bonds to hydrogen atoms are kept fixed. Schematics of the polymers is shown in Fig. 1.

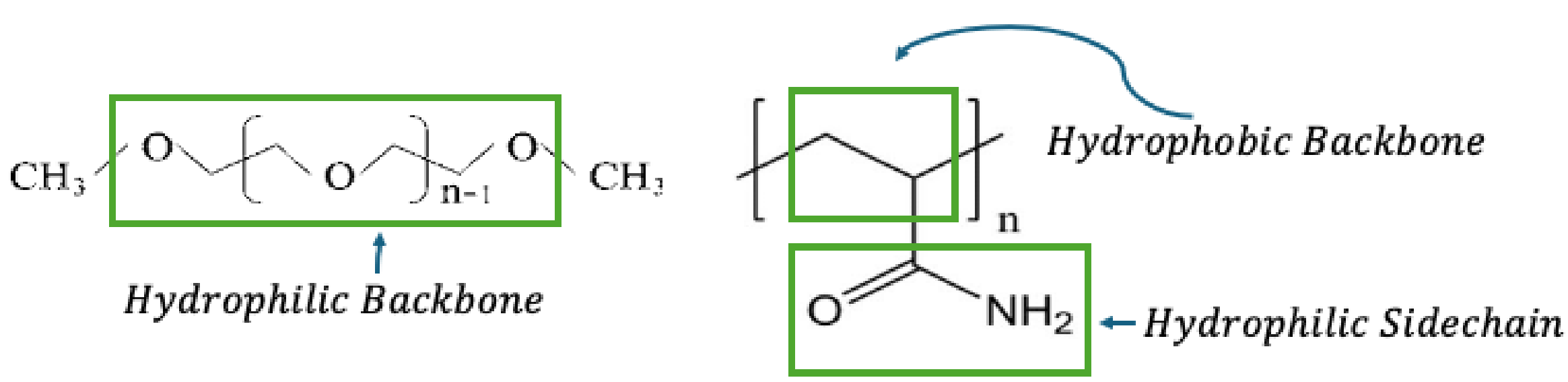

***Figure 1**: Schematic of the molecular structures of PEO (left) and PAM (right).*

We use the non-polarizable TIP4P/2005 model[37] to represent water molecules. Parameters for salt interactions ate taken from Li et. al. [24]. These parameters were originally developed for TIP4P/Ew model of water but based on a recent study[38] that shows excellent transferability of ion parameters between TIP4P/Ew and TIP4P/2005 water models, we adopt the parameter sets for further investigation. Monovalent $\mathrm{Na^+, K^+, Li^+, NH_4^+}$ and $\mathrm{Cs^+}$are are used as cations in this work, keeping the anion fixed as $\mathrm{Cl^-}$. The sequence of the cations on the HS according to their protein precipitation ability is[39] $\mathrm{NH_4^+ > Cs^+ > K^+ > Na^+ > Li^+}$, where $\mathrm{NH_4^+}$ and $\mathrm{Cs^+}$ can be classified as chaotropic, and $\mathrm{Na^+}$ and $\mathrm{Li^+}$ as kosmotropic ions, while $\mathrm{K^+}$ and $\mathrm{Cl^-}$ sit somewhere in the middle, showing little to no effect on the solubility. All salt and water atoms interact with each other and with polymer atoms via non-bonded pair potential given by Eq. (1) and cross interaction terms follow the geometric mixing rule. The salt parameters have been shown to reproduce the hydration free energy for the cations considered in this study.

For long-range corrections to the Coulombic interactions, we use the particle-particle-particle-mesh (PPPM) method. LJ interactions are truncated after a cut-off of 1.2 nm. The SHAKE[40] algorithm is used to constrain bond lengths and angles wherever required. Simulations are performed at a constant temperature 300 K (unless otherwise stated) and at a constant pressure (1 atm). To create initial configurations for the polymers, we have used Moltemplate package[41].

We carry out our molecular dynamics simulations using the Large-scale Atomic/Molecular Massively Parallel Simulator (LAMMPS) package[42]. For simulations of salt-water system, we use 5000 water molecules and a varying number of ion pairs to get desired concentration of salts. For simulations involving polymers, the simulation box contained a single 50-mer chain of either PEO or PAM. To hydrate the polymers, we use 29900 water molecules for PEO simulations and 22880 water molecules for PAM simulations. Periodic boundary conditions are applied to the simulation box. To study the effect of salts on polymer solutions, we add 350 pairs ions, resulting in a salt concentration of approximately 0.65M and 0.8M in PEO and PAM solutions respectively. Starting from random initial configurations, we equilibrate each system for 40 nanoseconds (ns) using an integration step size of 2 femtoseconds (fs); production runs use a smaller 1fs timestep for at least 70 ns duration.

We calculate the self-diffusion coefficient $D$ of the water molecules in different salt solutions from the mean-squared displacement $\langle r^2(t)\rangle$ (MSD) using the Einstein relation[43] $\langle \mathrm{r}^2(\mathrm{t})\rangle = \lim_{t\to\infty} 6\mathrm{Dt}$. For TIP4P/2005, the MSD data after ≈ 5 ps can be used to fit a line to extract $D$[44]. We have used two decades in time (10 ps to 1000 ps) to get an accurate estimate of D. Data for calculating D is taken from a 10 ns window during the production runs. To understand the picosecond (ps) dynamics of water in the hydration layer, we measure the Debye-Waller parameter, $\langle \mathrm{u}^2\rangle$, the average MSD at 1ps interval[45]. This quantity captures the local mobility of water molecules at the caging timescale. To understand how hydrogen bond networks are affected by the addition of salts, we employ a hydrogen bond lifetime (HBL) analysis. This time is calculated from the correlation function[46],

$$C_{HB} = \frac{\langle h(0)h(t)\rangle}{\langle h(0)\rangle}, \tag{2}$$

where $h(t)$ has the value 0 (no hydrogen bond) or 1 (hydrogen bond present). We fit the decay of $C_{HB}$ with time with two exponentials[47] to extract the lifetime $\tau_{HB}$, as a weighted sum of the two-time constants of the exponential functions[48]. To define the hydrogen bonds, we use a standard distance cut-off of 3.4 Å between donor and acceptor atoms. The angle cutoff[49] for hydrogen-donor and hydrogen-acceptor is set to $150°$ .

We evaluate the polymer radius of gyration, $R_g$, to quantify the chain size; we compare these values with the hydrodynamic radius, $R_h$, calculated using ZENO[50,51].

We quantify relaxation of water molecules using the self-part of the intermediate scattering function[52]

$$F(k,t) = \frac{1}{N}\left\langle \sum_{i=1}^{N} \exp[-\mathrm{i}\boldsymbol{k}\,.\{\mathbf{r}_\mathrm{i}(\mathrm{t}_0+\mathrm{t}) - \mathbf{r}_\mathrm{i}(\mathrm{t}_0)\}]\right\rangle, \quad (3)$$

where $k = |\boldsymbol{k}|$ is the wave number and $\boldsymbol{r}_i$ is the position of $i^{th}$ particle; $\mathrm{t}_0$ and $\mathrm{t}$ are strating time and waiting time respectively. $\langle ... \rangle$ denotes ensemble averaging.

We characterize the local structure of the water molecules by using the tetrahedral order parameter[53]

$$q = 1 - \frac{3}{8}\sum_{i=1}^{3}\sum_{j=i+1}^{4}\left(\cos\psi_{jk} - \frac{1}{3}\right)^2, \quad (4)$$

where $\psi_{jk}$ is the angle defined by the nearest neighbors $j$ and $k$. The tetrahedral order parameter varies over the range $0 \le q \le 1$, where 0 marks the ideal gas (random) orientation and 1 corresponds to a regular tetrahedron[54].

## III. Results

### (A) Dynamics and Structure of aqueous electrolyte solutions:

Before simulating aqueous polymer solutions, we first need to establish the ability of the model to correctly capture the specific ion effects on the dynamics of salt solutions. This is vital given the strong correlation of ion effects to the properties of the polymer solution, especially ion mobility trends. Because earlier works have shown the difficulties capturing the qualitative trends of ions on water mobility[11,27,55], we start with the effect of ions on the water diffusion constant. Fig. 2 shows the salt concentration dependence of the self-diffusion coefficient of water relative to $D_0$, the value of the quantity in salt-free TIP4P/2005 water model. The concentration dependence seen qualitatively matches the experimental trends for these ions[56] where chaotropic salts are found to enhance the self-diffusion constant of water, while kosmotropic ions reduce the relative magnitude water diffusion coefficient. The effect of KCl on the self-diffusion coefficient is rather small, in line with existing experimental data,[57] and for the remainder of this paper, we will focus on the effects of the remaining four salts. The fact that our results are qualitatively in-line with experiments provide us the confidence to proceed further. We continue with further characterization of the effects of ions on water before considering the addition of polymers.

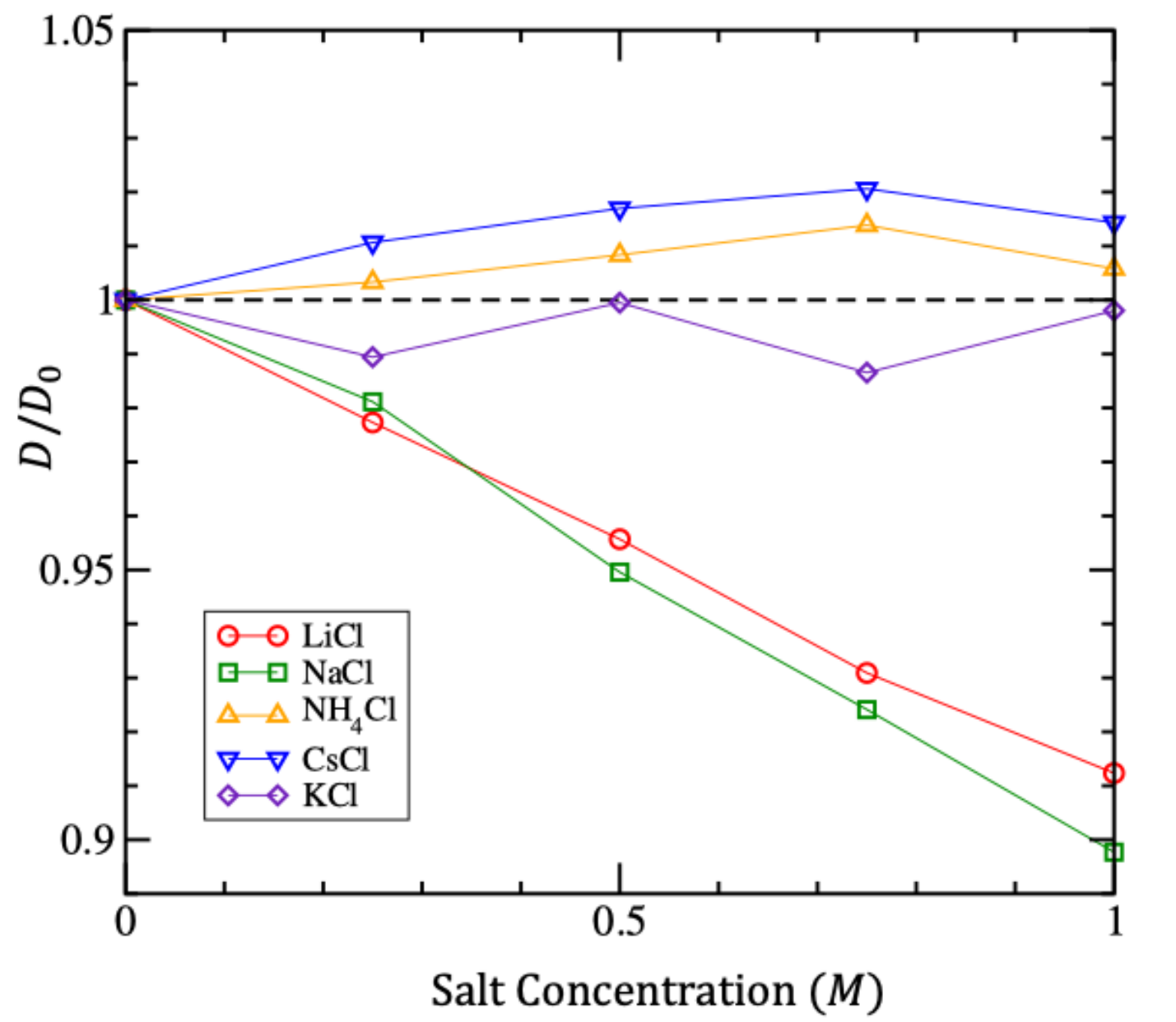

***Figure 2**: Water self-diffusion constant, $D$, in electrolyte solution, relative to its value in pure bulk water system, is plotted against concentration (Molarity) of salts. Data for five different salt types are presented.*

We next characterize the structure of the ion pairs for chaotropic and kosmotropic ions. Fig 3(a) shows cation-$Cl^-$ pair distribution function $g(r)$ in different salt solutions at 0.5M salt concentration. Li shows a very sharp and intense peak around 2.4 Å followed by Na, $NH_4$ and Cs. These peaks correspond to contact ion pairs (CIP), i.e., ions without intervening water molecules. $Li^+$ and $Na^+$ also exhibit a second peak, shown in the inset, between 4 Å and 6 Å, which corresponds to ion pairs that are mediated by a layer of water molecules (SMIP). As the size of the cation increases, separation between ions increase and both CIP and SMIP become less stable. We see no discernable second peak for $NH_4$ and Cs, indicating that ion pairing only occurs directly in these cases. To illustrate this feature, Fig. 3(b, c) show snapshots of representative $Cs^+$, and $Li^+$ions forming ion pairs with common anion $Cl^-$, which only weakly hydrates. For CsCl, Fig. (3b) shows a a contact ion pair with surrounding hydration layer and Fig. 3(c) illustrates a solvent-mediated ion pair of LiCl. The scale of the hydration layer can also help us understand the amplitude of the changes to the water diffusion coefficients. For kosmotropic ions like $Na^+$ and $Li^+$, the extended hydration layer means that more water molecules are impacted by the ions, and correspondingly we see a larger amplitude change in $D$ than seen for the chaotropic ions like $NH_4^+$ and $Cs^+$ which have a smaller hydration shell.

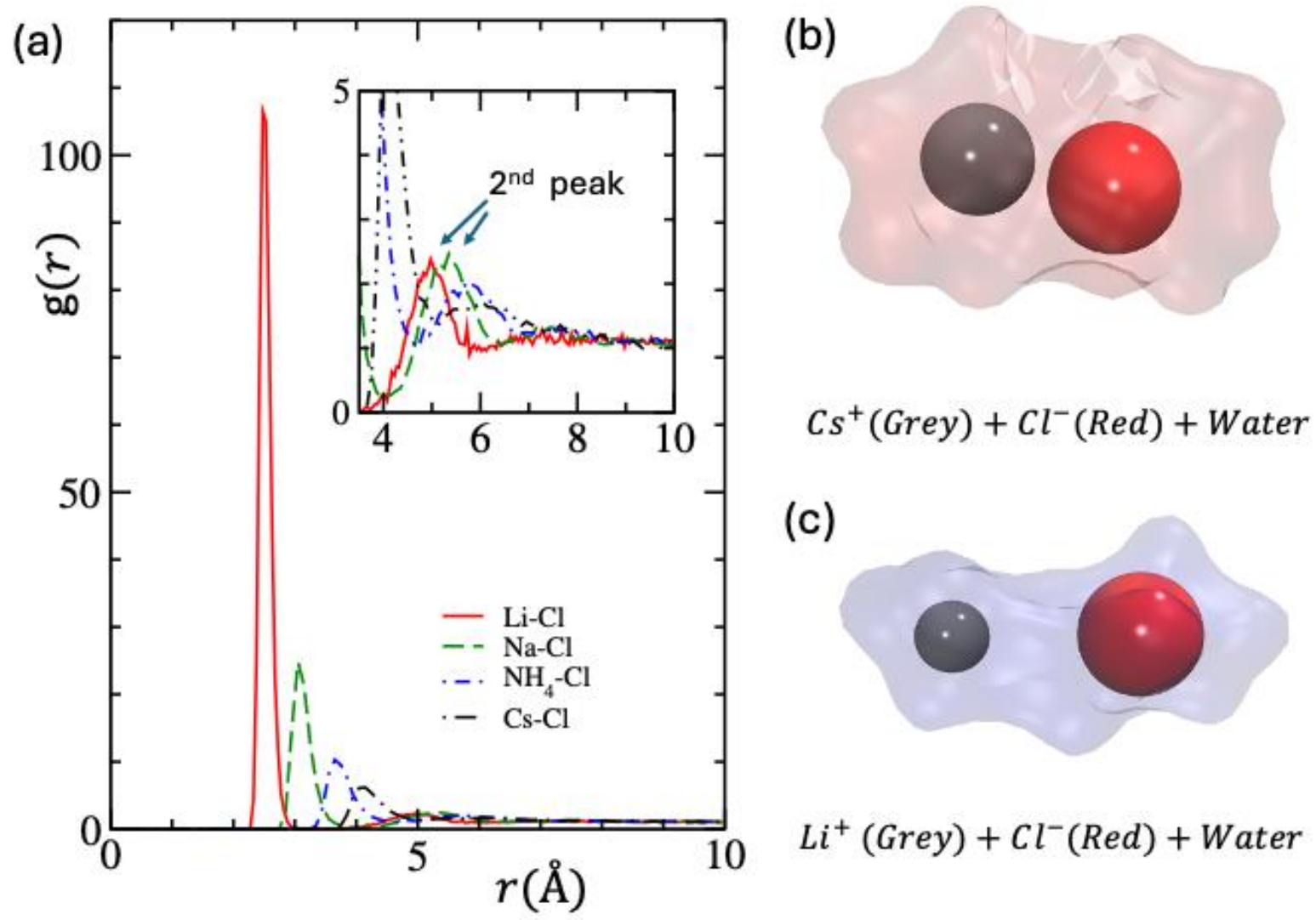

***Figure 3**: (a) Radial distribution function, g(r), for cation-halide pairs in different salt solution (0.5M) is shown. Inset shows a zoomed in version of the same to show the existence of a second peak. (a) Structural hydration layer (shaded region) around the contact ion pair of $Cs^+ - Cl^-$(c) The structural hydration layer for second neighbor $Li^+ - Cl^-$where pairing is mediated by the hydration water.*

We now proceed to examine the dynamic behavior of water as a function of distance from the cation. To quantify the dynamics of water molecules in the vicinity of ions (and later polymers), we evaluate the distance dependence of the Debye-Waller (DW) parameter, $\langle u^2 \rangle$, defined earlier. Although $\langle u^2 \rangle$ is estimated on a ps timescale, this quantity has been shown to be predictive of mobility on much longer timescales; this relationship has been discussed for a wide variety of glass-forming liquids[58–60] as well as in water simulations[60]. We discussed in detail the significance of this quantity and its relationship with the free volume for TIP4P/2005 water model in the supporting information. Fig. 4 shows $\langle u^2 \rangle$ calculated for various salt types at different salt concentrations. Fig. 4(a) shows a color map of this quantity as a function of distance from the cations showing more and less intense changes of mobility. Fig. 4(b) plots $\langle u^2 \rangle$ as a function of distance from the cation, relative to its value $\langle u_0^2 \rangle$ in pure water. Unexpectedly, in *all* cases, water mobility near the ion interface is reduced, albeit to different extents, even for those salts ($NH_4Cl$ and CsCl) where the overall diffusion coefficient *increases* with salt concentration. To understand how this is possible, we recognize that, for these salts, further from the ion, $\langle u^2 \rangle$ actually increases relative to pure water. For $Li^+$ and $Na^+$, where the overall diffusion is strongly reduced, we observe a corresponding significant reduction in local water mobility close to those ions and the mobility tends towards the salt-free limit at large distances. It is also interesting to note that the amplitude of the change in mobility around various ion types is nearly independent of the salt concentration. For reference, Fig S3 in the supporting information shows $\langle u^2 \rangle$ of water molecules around the chloride ions present in different salt solutions.

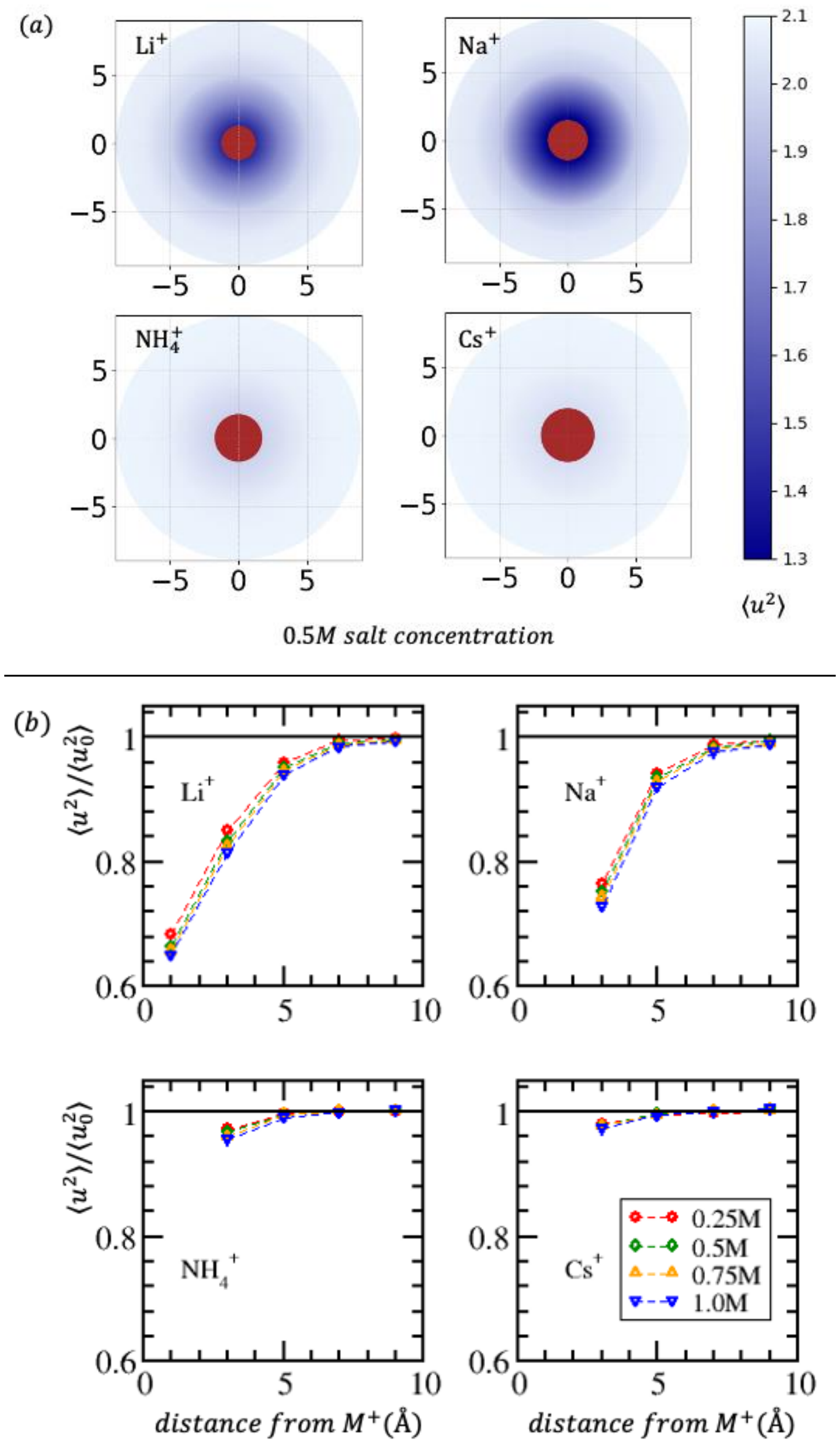

***Figure 4:*** *Debye-Waller parameter, $\langle u^2 \rangle$, around different cations in dilute electrolyte solution (a) corresponds to a salt concentration $0.5M$. (b) shows the spatial variation of $\langle u^2 \rangle$ for the water molecule about the ion, relative to its value in salt-free case $\langle u^2(0) \rangle$, is shown as a function of distance from the cations ($M^+$) present in the simulation box, for different salt concentrations.*

### (B) Quantifying the Effects of Salts on the Dynamic Hydration of Representative Polymers

It is well known that the properties of water surrounding the polymer chain differ from those of bulk water[61,62]. The extent of the first or the static hydration layer is typically defined by the distance of first minima in the pair-correlation function between hydrophilic species on chain and water. This is the layer where water molecules are in direct contact with the chain. Fig. 5 shows snapshots of equilibrium configurations of PAM and PEO chains along with their static hydration shells. These results are in qualitative agreement with earlier simulations of short chains.[6,63,64]

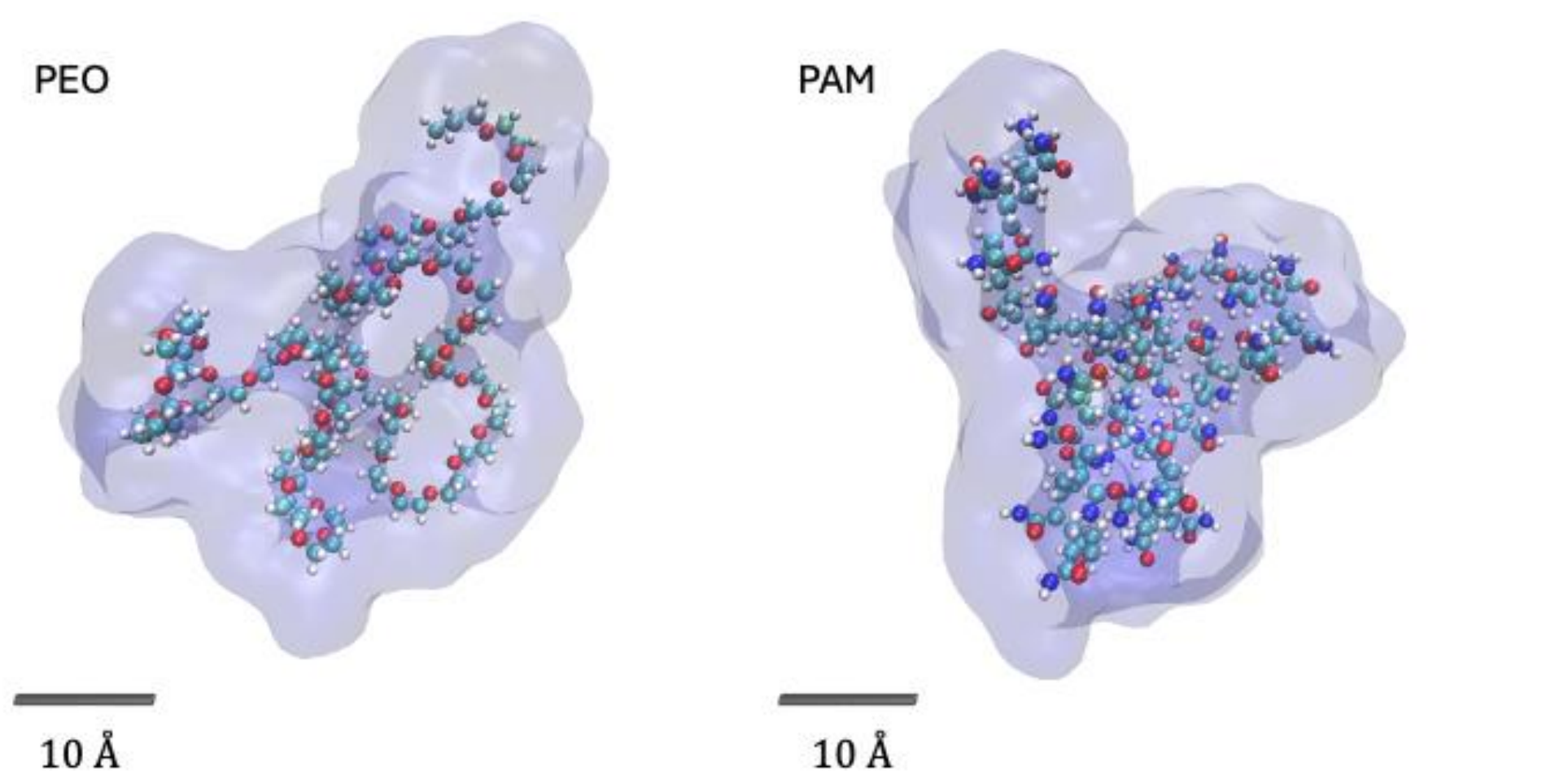

***Figure 5****: Snapshot of PEO (left) and PAM (right) configurations at equilibrium obtained from simulations in water with no added salts. Oxygen, Hydrogen, Nitrogen and Carbon atoms are shown in red, yellow, blue and orange colors respectively.*

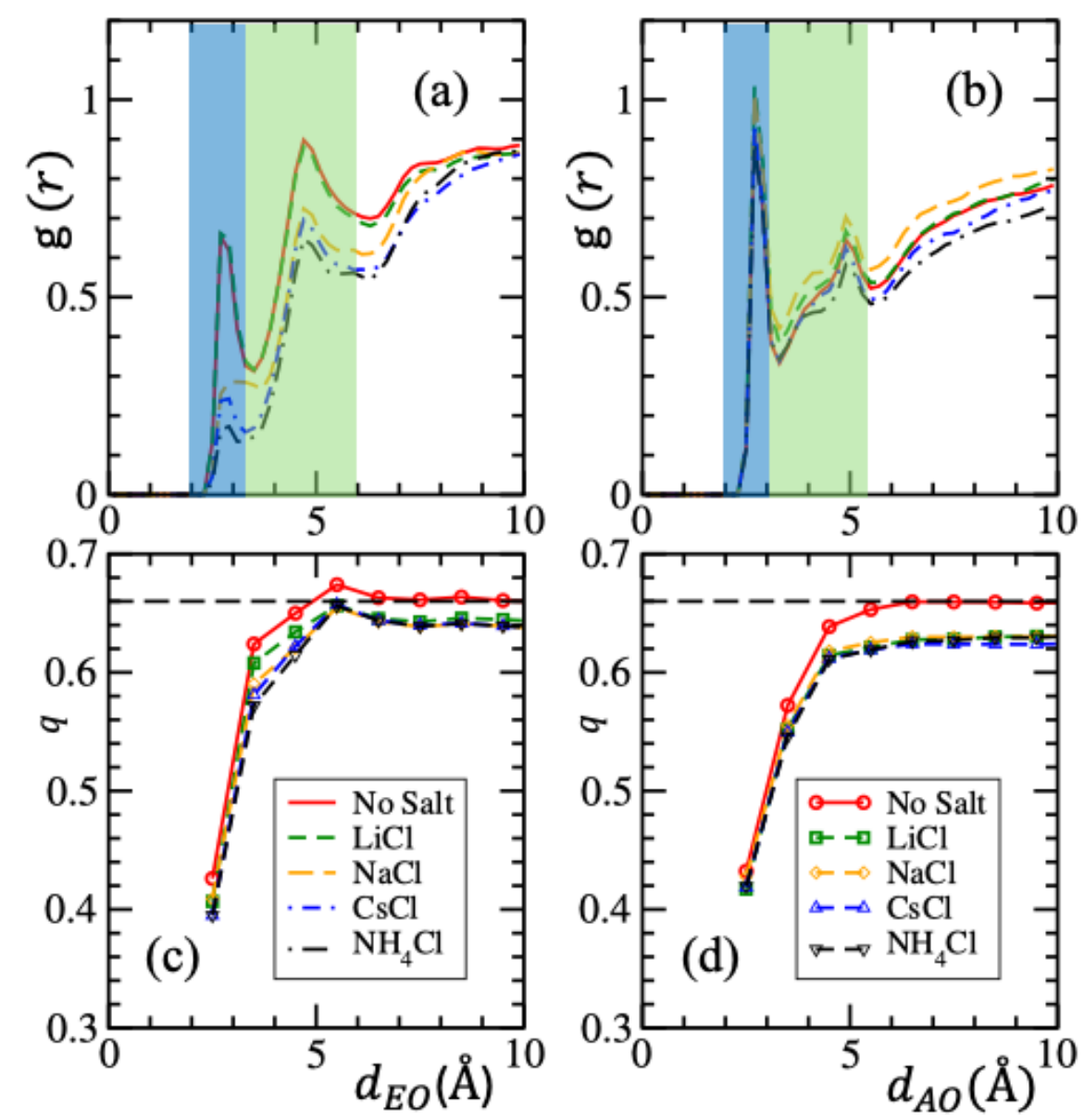

***Figure 6**: Figures (a) and (b) shows the radial distribution function of water Oxygen and EO/AO respectively. The extent of first and second hydration layers are highlighted by blue and green shades respectively. (c) and (d): the tetrahedral order parameter, $q$, of water is plotted versus distance, $d$, measured from the EO and AO sites on PEO and PAM chains.*

We now proceed to quantify these properties of this hydration layers. A standard practice is to evaluate the changes in local order and density of water molecules. Fig. 6(a, b) shows the radial distribution function of EO-Water (Oxygen) and AO-Water (Oxygen) radial distribution function. The extent of first (sometimes called the *static* hydration layer*)* layer is defined by the first and second minimum of $g(r)$. Beyond the static hydration layer there is the dynamic hydration layer. The local density of water in the static hydration layer of PEO is significantly lower than that for PAM; different salts alter the density to different extent. For PAM chains, the water density goes slightly increases or decreases depending on the salt type. Another probe of the local structure is the tetrahedral order parameter, $q$, which in bulk water for TIP4P/2005 model[54] is $q \approx 0.66$, at room temperature. Nanoscale water-polymer interaction in the hydration significantly modifies the local tetrahedral order. Fig. 6(c, d) shows a comparison of $q$ in in PEO and PAM in different salt solutions as function of distances from the hydrophilic sites present on the polymer. For both PEO and PAM, tetrahedral structure is disrupted inside the *first* or the static hydration layer, shaded with blue on $g(r)$ plot. Beyond this layer, for PAM, the polymer-water interaction seems to have a very little effect on the bulk water structure. On the other hand, for PEO, there is a small enhancement in tetrahedral structure beyond the static hydration layer (between 3.5 Å and 6 Å), demonstrating that these polymers alter the structure of the solvent, subtly at somewhat larger distances than the static hydration layer defined above. Nonetheless, the structure defined in terms of local density is not appreciably beyond the static hydration layer and even topological measures of local order extend out only a small distance beyond the static hydration layer.

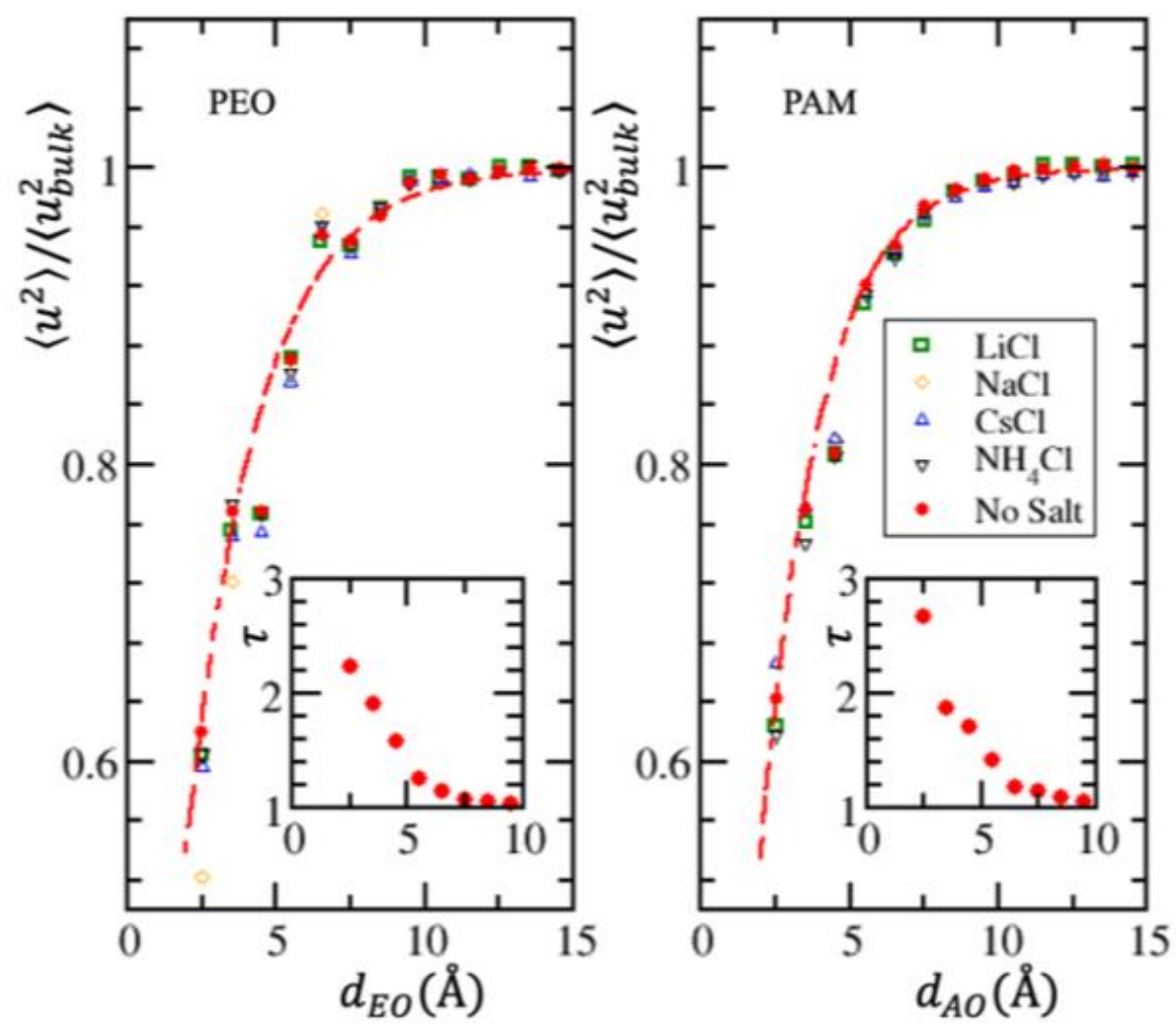

***Figure 7**: Debye-Waller parameter, $\langle u^2\rangle$ , normalized by its value at bulk, is plotted as a function of distance from the electronegative site ($N^-$) on the polymer chains. The dashed line is a fit to data for system with no salt, using Eq. (5). Inset shows the relaxation time of water molecules as a function of distance from the polymer surface, relative to its value at bulk.*

We expect that the structural changes near the polymer chains are accompanied by changes in local dynamics. We quantify the mobility of water molecules in the vicinity of the polymer chain by again evaluating $\langle u^2\rangle$ as a function of distance fro the chain; specifically, Fig. 7 shows $\langle u^2\rangle$ relative to bulk water molecules (those located $> 20$Å away from the chain) as a function of distance from EO/AO sites. A previous simulation work, involving study of a single polyzwitterion in dilute solution showed the presence of an extended dynamic hydration layer[45]. A variety of experimental methods[65–68] have indicated the existence of such a dynamic hydration layer extending out to a distance on the order of a $nm$ in proteins and DNA, while the scale tends to be somewhat reduced in water-soluble small molecules such as sugar molecules[69]. Evidently, the existence of an extended dynamic hydration layer of reduced mobility relative to bulk solvent seems to be universal. Experimental measurements have not yet been able to resolve the mobility gradient within the dynamic hydration layer seen in our simulations as this requires greater measurement resolutions.

We can quantify the range of these dynamic changes by fitting the distance dependence of $\langle u^2\rangle$ with an exponential function[70],

$$\langle u^2\rangle/\langle u^2_{bulk}\rangle = 1 - \delta u \exp\left(-\frac{d}{\xi}\right), \tag{5}$$

where $\delta u$ is the difference from the saturation value, $d$ is the distance from the electronegative site on the chain and $\xi$ is the measure of the range of the changes in $\langle u^2\rangle$. The fitting of Eq. 5 to our $\langle u^2\rangle$ data in Fig. 7 is shown by dashed lines. The interfacial width is found to be slightly smaller for PAM compared to PEO, but added salts seems to have little to no effect on the local mobility of water (when scaled by the bulk mobility). For completeness, the inset of Fig. 7 shows that the relaxation time ($\tau$) of the intermediate scattering function of water molecules conditioned on their distance from the polymer mimics the behavior of $\langle u^2 \rangle$. Considering that we observe a substantial reduction in tetrahedral order parameter in all cases, we might have anticipated an increase local in mobility. Apparently, the relationship between changes in local order and dynamics are more complex than this naïve view.

We also analyze the dynamics of hydrogen bonds to understand the differences in the first hydration layers around the hydrophilic sites of our model polymers and the effect of salts on these hydrogen bonds. Fig. 8 shows the characteristic hydrogen bond decorrelation time obtained from the correlation function defined by Eq. 2 for bonds formed between EO/AO site on the chain and water molecules. We find that the hydrogen bonds between oxygen and water in static hydration layer of PEO chain have much shorter lifetime than those formed between the amide oxygen on PAM and water. The effect of salts on the hydrogen-bonding lifetime does not appear significant.

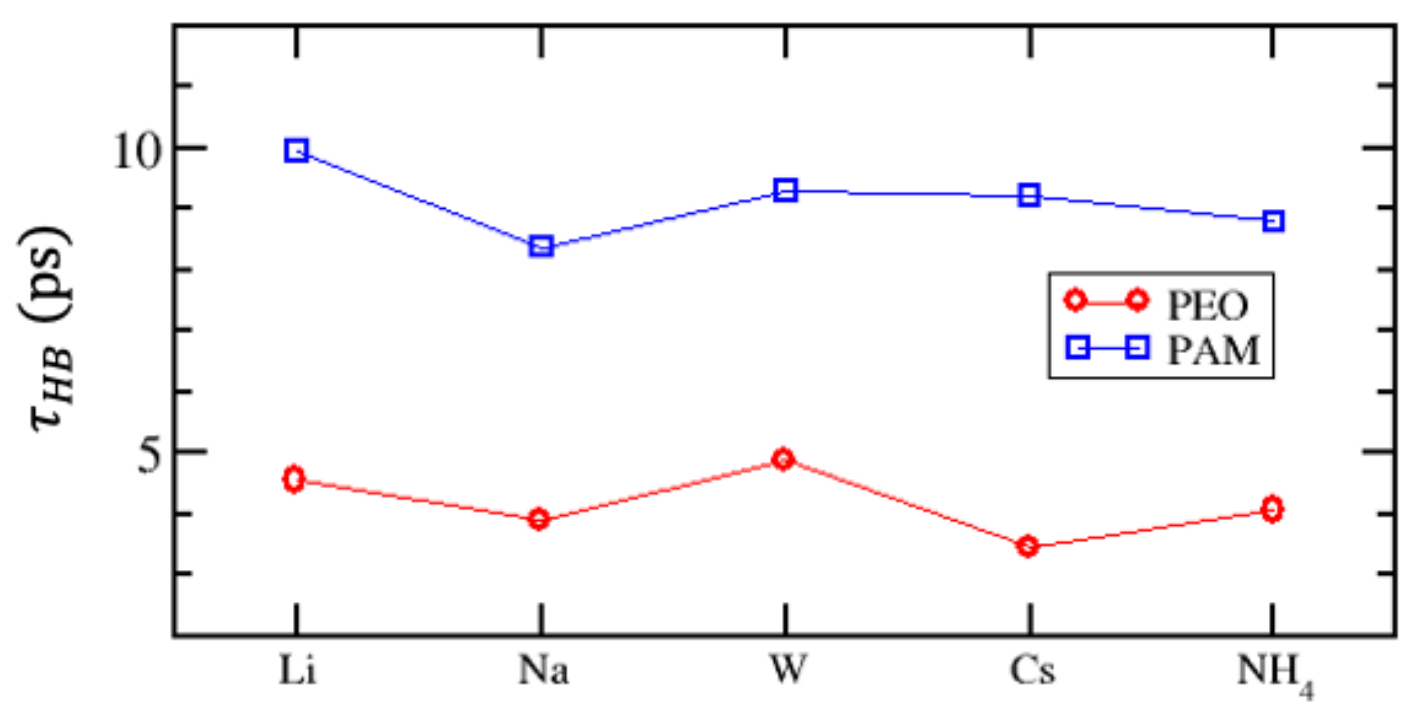

***Figure 8**: Intermittent lifetime, $\tau_{HB}$, calculated between PEO/PAM and water molecules in static hydration layer is shown.*

We investigate the adsorption of cations onto the electronegative sites of the polymer chains, the ether-oxygen (EO) site of the PEO and amide oxygen (AO) site in the carbonyl group of the PAM chain, via the running coordination number,[71]

$$n(r) = 4\pi\rho \int_0^r r'^2\, g(r')dr', \tag{6}$$

where $r'$ is the separation between the ion and charged site of the polymer and $g(r')$ is the radial distribution function[35]. Fig. 9(a) shows $n(r)$ of EO-cation in saline solutions indicating a trend $NH_4^+ > Cs^+ > Na^+ \gg Li^+$, where \` $>$ ′ implies a higher condensation. Fig. 9(b) shows similar quantities but for the PAM solutions. The ordering of the cations is this case is found to follow the same sequence as the case of the PEO solutions, while the overall adsorption for any cation is found to be much less.

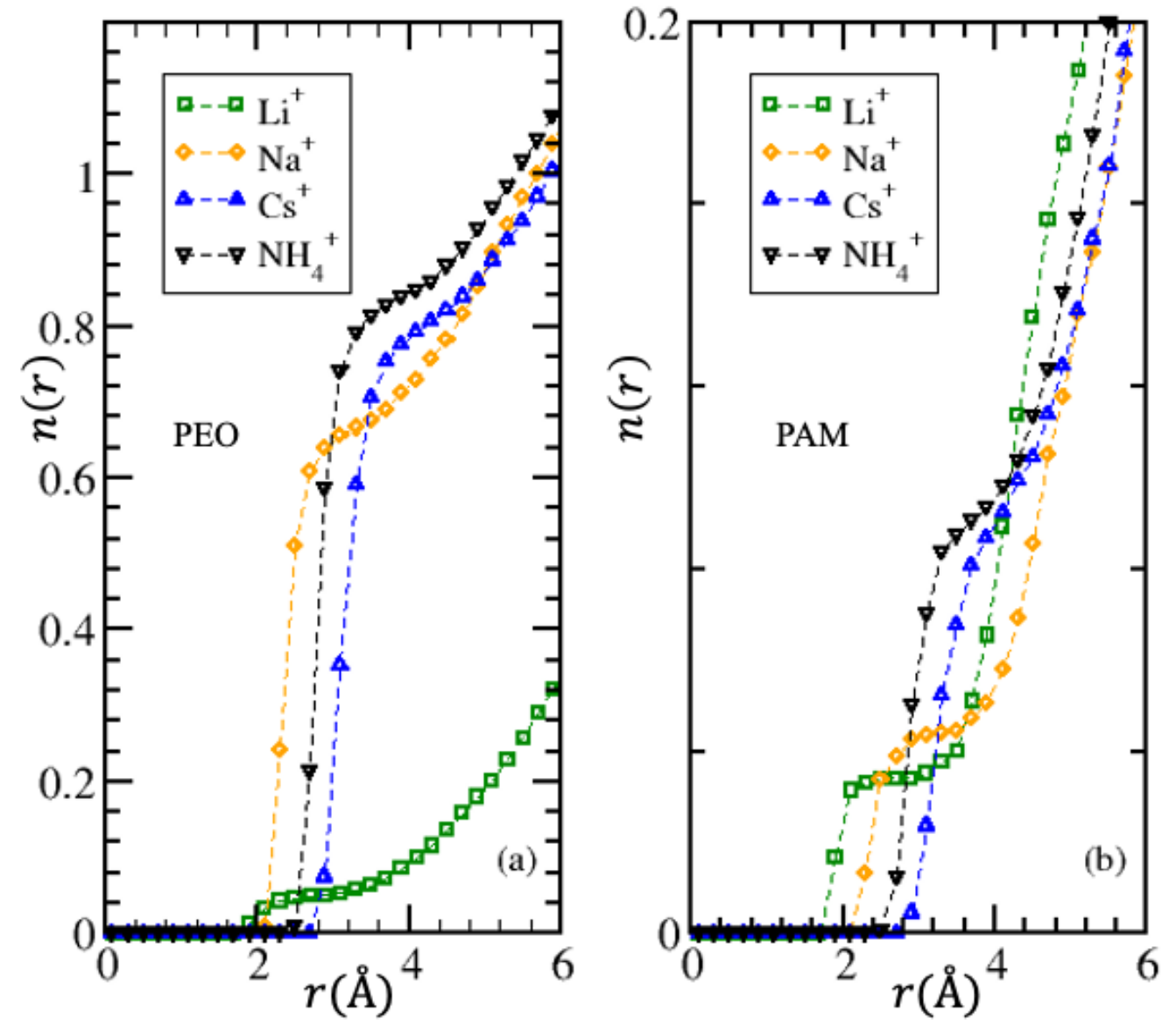

***Figure 9****: Running coordination number, $n(r)$ is plotted as a function of cations from electronegative sites on (a) PEO solution with salt concentration $0.65M$, and (b) PAM solution with $0.8M$ salt concentration.*

**(C) Change in structure of the chains due to added salts**

To quantify the changes in the chain size due to the salts, we evaluate the radius of gyration[64], $R_g$. Independent of the identity of the added salts, PAM shows a highly coiled structure; quantitatively, $R_g$ is slightly larger for PEO than PAM chain (Fig. 10). A related quantity that is readily measured experimentally is the hydrodynamic radius, $R_h$, which describes the radius of an equivalent sphere diffusing through the solution. We show both $R_g$ and $R_h$ values of PEO and PAM chains in salt solutions containing different cations as well as salt-free case in Fig. 10(a, b).

While, both $R_g$ and $R_h$ are used to gain insight into the structural properties of polymer chains, the ratio of these two quantities[13], defines the hydrodynamic penetration function $\psi_h = R_h/R_g$ [72,73], which provides information about the overall shape and 'compactness' of the macromolecule[72,74]. For example, the value of $\psi_h$ for a spherical particle[75] is $(5/3)^{1/2} \approx 1.29$ while for long rods, $\psi_h$ progressively approaches 0 with increasing length. The case of ideal random coil polymers, appropriate for highly flexible chains in solution, provides an intermediate situation. For polymers having configurations similar to a random walk, $\psi_h \approx 0.8$[13,76–78]. Note that $\psi_h$ is typically around 1 for branched polymer structures, such as lattice animals with repulsive, excluded volume interactions and percolation clusters (with screened excluded volume interactions), polymer grafted nanoparticles[79], "tumbleweed" morphologies of carbon nanotubes[80] and many-arm star polymers[81,82]. The value of $\psi_h$ can also depend somewhat on the degree of polymerization[83].

Notably, the accurate calculation of $\psi_h$ by analytic theory without the introduction of crude mean field "preaveraging" approximation is difficult because of the large fluctuations in conformational shape of random coils that cause large fluctuations in $R_g$.[78,84] Fortunately, the ZENO computational method allows for this difficulty to be overcome. Generally, when a molecule adopts a more extended shape, $\psi_h$ decreases, such as increasing the strength of the repulsive polymer excluded volume interaction in flexible linear polymers or making the polymer relatively stiffer. Evidently, the addition of salt has a greater influence on polymer average shape in PAM than for PEO, but both polymers exhibit more compact configurations than random coil polymers.

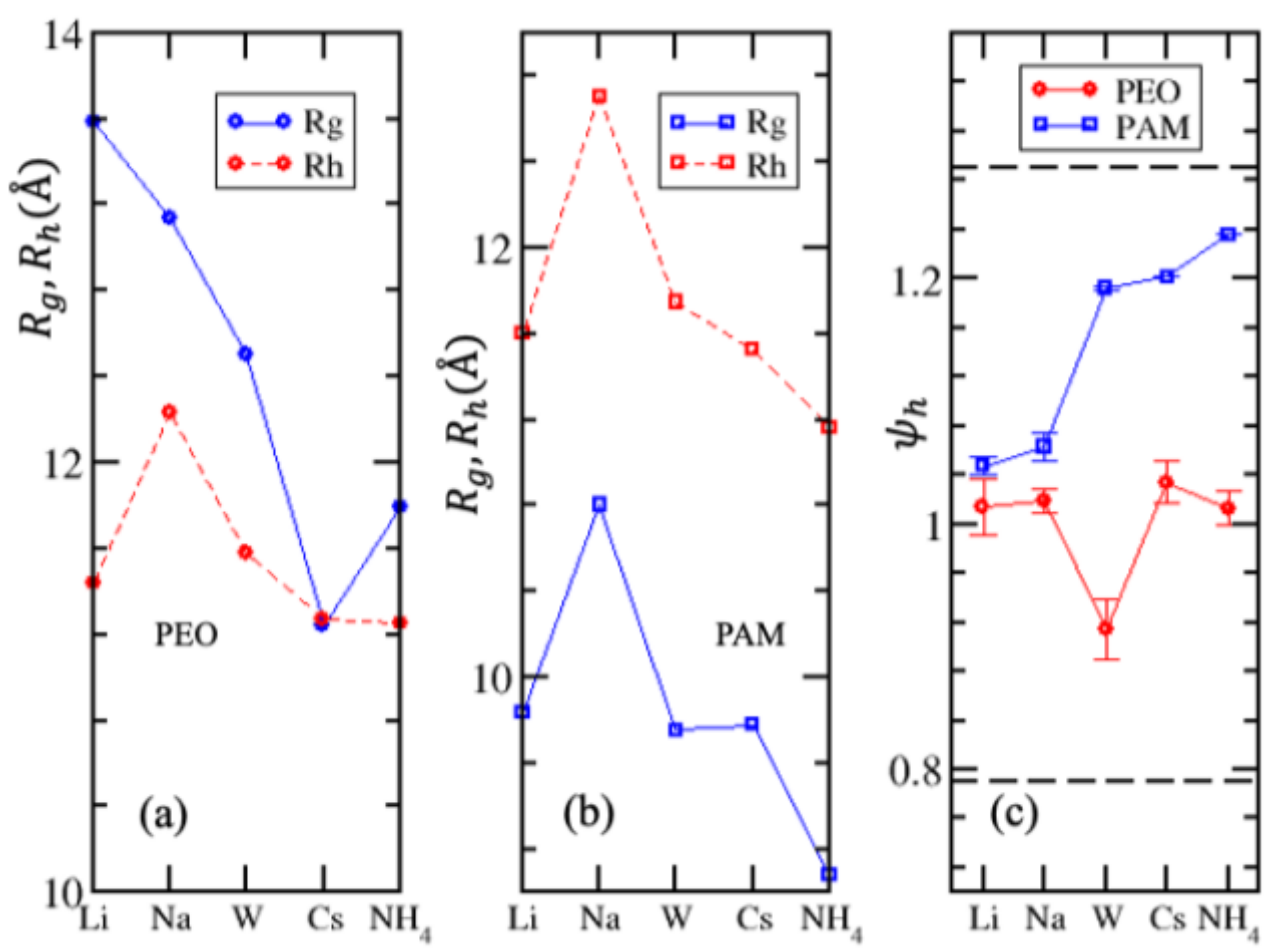

***Figure 10**: Average radius of gyration, $R_g$, and average hydrodynamic radius, $R_h$, are shown for (a) PEO and (b) PAM, for different salt solutions as well as aqueous solution without salts (marked with W). (c) Ratio of hydrodynamic radius and radius of gyration for both PEO and PAM in different salt solutions. The dashed lines indicate the values of this ratio for two different standard conformations of a linear chain. See text for details.*

In Fig. 10(c), we have plotted $\psi_h$ for both PEO and PAM chains in different salt solutions, with mean squared deviations as error bars. Evidently, the addition of salt has a greater influence on polymer average shape in PAM than for PEO, but both polymers exhibit more compact configurations than random coil polymers. The $\psi_h$ data show a weak dependence of PAM shape on salt type while PEO chains show no such effect. We observe a slightly swollen PAM chain in the presence of kosmotropic ions like $Li^+$ and $Na^+$. On the other hand, addition of chaotropic ions lead to a highly compact structure of PAM chains; PAM adopts a nearly globular shape in the presence of $NH_4Cl$. This trend in conformational change of the PAM chain is consistent with the Hofmeister series[21], where the ions are arranged in a sequence depending on their ability to act as a salting-out agent to a salting-in agent. The Hofmeister series seen in protein solutions is a notoriously complex phenomenon. It evidently involves selective association of hydrated ion species with the hydrated protein or other biopolymer, altering the propensity of macromolecules to associate with each other through the alteration of both the thermodynamic interaction strength mediated by the biomolecule hydration layer and the polymer rigidity: these factors acting together to alter the strength of the molecular binding interaction. Of course, the details of these coupled association processes remain to be investigated to better understand the Hoffmeister series from a more fundamental perspective. Nonetheless, the degree of ion hydration should play a large role in the binding process between biopolymers, so it does not surprise us that a 'correlation' exists between the magnitude of degree of hydration, quantified by the magnitude of the viscosity $B$ coefficient, and the Homeister series. That said, other factors, such as the hydrophobicity of the ions and polymer, the charge on the chains and thus pH, etc. must also influence the association of ions to the polymer and the Hofmeister series. So, this correlation merely indicates a trend that can be useful in applications rather than any kind of quantitative relationship.

## IV. Discussion and Conclusions

The realistic modelling of the effects of added salts on the dynamics of aqueous polymeric solutions by classical molecular dynamics simulation is a long-standing problem of concern because of the correlation of ion-specific effects with the Hofmeister series and other basic properties of aqueous solutions, such as self-diffusion coefficient and shear viscosity. We make a modest attempt to improve our understanding of this type of system starting from a reasonably good potential model of water, TIP4P /2005, and potential parameters for ions parametrized based on the calculation of hydration free energies. We investigated the spatial extent of the hydration layers around monovalent metal ions, in electrolyte solutions, ranging from chaotropic to kosmotropic, with a fixed anion. In addition to reproducing the general trends in ion-specific effects on water mobility without introducing ad hoc assumptions, such as arbitrary choices of partial charges considered in previous modelling, we find that strongly hydrating kosmotropic ions have a highly extended hydration layer that extends out to a scale on the order of one nm. Water molecules within the hydration layer of such ions are found to have a significantly lower mobility than in salt-free case; this "dressing" of the kosmotropic ions influences the nature of ion pairing. We have found solvent mediated ions pairs in solutions of kosmotropic ions that are absent in solutions containing chaotropic ions.

After that we can reproduce the basic ion-specific mobility series within the TIP$P/2005 water model without introducing unwarranted approximations, we quantified the effect of these ions on hydration of model water-soluble polymers. The polymers investigated, PEO and PAM represent two different class of neutral polymers. PEO is a polyether, with a $C-O-C$ backbone which is hydrophilic and no side chains. On contrast, PAM is a vinyl polymer, with $C-C$ hydrophobic backbone and a short side chain with strong hydrophilic amide group. The hydration layer is a layer of water wrapping the polymer, whose properties differ from those in the bulk. A key concern is the exact spatial extent of this hydration shell. Many different techniques[68] have been used in experiments to gain insight about the hydration shell dynamics. NMR spectroscopy[85], small-angle x-ray and neutron scattering[86], and IR spectroscopy[87] experiments directly probe the dynamics of individual atoms and in dilute solutions, the collected data is usually dominated by bulk water. Quasi-elastic neutron scattering (QENS) experiments[88] are used in study of biomacromolecules with high molecular weight. The development of the terahertz spectroscopy[89] method was a major leap in understanding the hydration layers as it showed that for macromolecules, the hydration shell extends far beyond the static hydration shell of adjacent water molecules that are effectively bound to the polymer. Here, we have used the local mobility, $\langle u^2 \rangle$, average mean-squared displacement of the water molecules, calculated at a fast $\beta$-relaxation timescale (1ps), as an efficient way to quantify the hydration layer around the polymer chains. This quantity is closely

related to the $\alpha$-relaxation process and hence related to the structural relaxation and diffusion. We investigated this connection between the $\alpha$-relaxation time and $\langle u^2 \rangle$ for the TIP4P/2005 model of water and presented it in the Supplementary Information. Inside the static hydration layer, extending up to $\approx 3.2$ Å from the hydrophilic site, we can see a $40$ % reduction in $\langle u^2 \rangle$, relative to its value in bulk. With increasing distance from the surface, $\langle u^2 \rangle$ increases exponentially to ultimately reach a bulk value marking the extent of the hydration layer; approximately $1$ nm. Importantly, this scale appears insensitive to the polymer or salt type. The outcome on the extent of the hydration layer is remarkably similar to results obtained earlier for many different proteins[90] and confirms an apparent universal trend in dynamic hydration of polymeric systems. The functional form for the distance dependence $\langle u^2 \rangle$ appears also to be insensitive to the type of polymer and ion identity. The structure of the water molecules provides little insight into this mobility gradient around the polymers. The addition of salts to the polymer solution changes the mobility in the hydration layer but also changes the water mobility in the bulk solution in such a way that the gradient found in the absence of salts remain unchanged, within numerical uncertainty. In contrast, the variation in water structure and density profile inside this extended hydration shows a nonuniversal ion-specific and polymer-specific trend.

We also analyzed the overall shapes of the polymers with and without salts with a view toward identifying specific ion effects. The shape and size of the polymer chains are found to depend slightly on the type of salt present in the system. For PEO, the size of the chain increases slightly in the presence of kosmotropic salts, while for chaotropic salts, the chain is found to contract somewhat. In general, the chain takes a more globular shape in presence of salts than a chain in the salt-free solution, irrespective of the ion type. For PAM, the salt effect on the size of the chain is rather complex. When we look at the shape, or the compactness of the chain configuration, we find a trend in PAM chains that closely follows the Hofmeister series. We further found that adsorption of cations on the electronegative oxygen sites on the polymer chains follows the order $NH_4^+ > Cs^+ > Na^+ > Li^+$, which leads us to beleave that the adsorption of ions is strongly correlated to dehydration and precipitation of the polymers.

The present work, as well as most measurements are limited to a constant temperature of $300$ K (near room temperature). In the future, it would be interesting to study polymer solutions over a range of temperatures both in presence and absence of salts to determine effects studied in this paper. Chaotropic and kosmotropic salts have been observed to influence water mobility differently in confinement[91,92] than in bulk system. This would be another topic to study experimentally and computationally since significant changes in the hydration layer of confining boundaries and polymers in confined environments can be expected.

**Supplementary Material**

The supplementary material of this article contains plots of self-intermediate scattering function of pure water at various temperatures and plots connecting the $\alpha$-relaxation time to the picosecond-timescale dynamics for the TIP4P/2005 model. We have discussed the relation between these two from the dynamic free volume perspective. It also includes a plot of the DW parameter of water around the $Cl^-$ ions in electrolyte solution.

**Acknowledgements**

Most of the computations were performed using the central high-performance computing facility at Wesleyan University. Post processing and system preparations involved systems provided by National Institute of Standards and Technology (NIST). The authors acknowledge funding from NIST award 70NANB24H025.